# Studies of Exoplanets with Candidate TOI 717.01 and Confirmed HAT-P-3b

*Sujay Nair*
*Stanford Online High School*
*Academy Hall Floor 2 8853, 415 Broadway Redwood City, CA 94063*
*sujaynr@ohs.stanford.edu*

*Krithi Koodli*
*Stanford Online High School*
*Academy Hall Floor 2 8853, 415 Broadway Redwood City, CA 94063*
*krithik@ohs.stanford.edu*

*Elliott Chalcraft*
*Stanford Online High School*
*Academy Hall Floor 2 8853, 415 Broadway Redwood City, CA 94063*
*elliottchalcraft@gmail.com*

*Kalée Tock*
*Stanford Online High School*
*Academy Hall Floor 2 8853, 415 Broadway Redwood City, CA 94063 kaleeg@stanford.edu*

**Abstract**

Images of the exoplanets TOI 717.01, Qatar-8 b, and HAT-P-3 b were requested in the w-filter from the Las Cumbres Observatory 0.4-meter telescopes. Of these requests, images were taken of TOI 717.01 and HAT-P-3 b and were reduced using the EXOplanet Transit Interpretation Code (EXOTIC) software, as well as 6 photometric algorithms from the Our Solar Siblings pipeline. HAT-P-3 b's transit midpoint was found to be 2458907.6205, which is 20.2 minutes different from its expected midpoint that night. For TOI 717.01, the transit was not discernible, likely due to its low transit depth of 0.1%. It is possible that one of the comparison stars in the TOI 717.01 field is variable.

## 1. Introduction

Exoplanets are planets that orbit stars outside of our solar system, and many exoplanets have been detected using the transit method. This involves plotting the apparent brightness of the host star against time and looking for periodic decreases in that brightness which correspond to planet occultations. The uncertainty of a planet's mid-transit time increases with time since the last measurement due to propagation of the uncertainty of its period. Because of this, continually observing and updating exoplanets' mid-transit times, also known as freshening the midpoints, is needed in order for predicted transit times to remain accurate (Zellem et al., 2020).

## 2. Target Selection

To select suitable exoplanet candidates for this study, we looked for planets that had the highest likelihood of being observed by the James Webb Space Telescope, the ARIEL mission, and the Astro2020 mission. High priority targets for these missions have atmospheric and thermal properties that make them especially interesting. Without repeated freshening, the target planets' mid-transit times are in danger of becoming uncertain before launch. Zellem et al.'s depiction, reprinted in Figure 1, illustrates the extent to which this uncertainty might affect the ability of these missions to optimize telescope time allocation. In addition to examining priority targets for upcoming space missions, we also looked for possible targets in the NASA Exoplanet Archive that would be observable by the 0.4-m LCO telescopes to which we had access.

The planets that were visible during the spring were identified using the LCOGT's seasonal visibility checker, which shows whether the LCOGT's telescopes are able to image the target within a given time frame. Figure 2 gives an example of the visibility of two different targets.

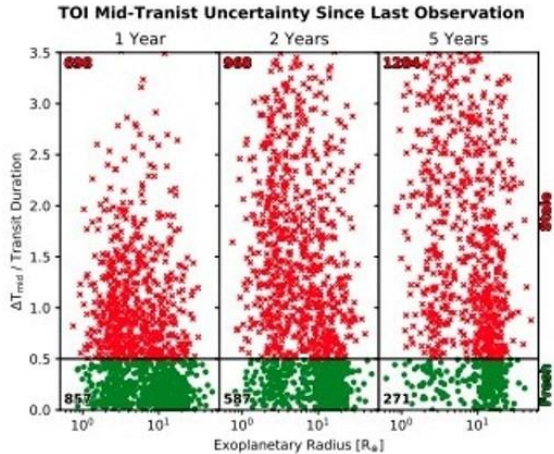

**Figure 1.** Uncertainty of mid-transit time increasing with the number of years since the most recent transit observation. Planets with smaller orbital radii and shorter periods tend to become stale more quickly (Zellem et al., 2020).

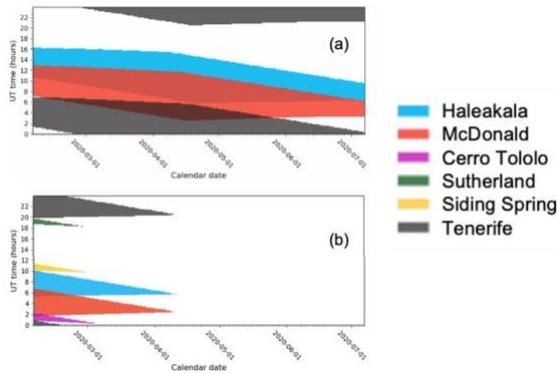

**Figure 2.** LCO visibility checker: (a) displays that the target HAT-P-3 b is visible throughout the season, whereas the exoplanet K2-155 b in (b) is only sporadically visible during the first part of March.

The selection of targets was further constrained to favor those with deep transit depths and short periods. A deep transit depth, usually of 1% or greater, is required for the LCOGT's 0.4-meter telescopes to see it. Additionally, short periods are necessary, as the project needed to be completed within a single semester, and it often takes multiple image requests before a complete transit is obtained. Because of this, the search was narrowed to planets with periods less than five days.

Based on these parameters, three targets were selected. Information on each host star and orbital parameters of the systems can be seen in Table 1 (at end of paper). TOI 717.01's transit depth was given as 1 millimagnitude (mmag) in the NASA Exoplanet Archive; we were initially unsure what this depth corresponded to in terms of a depth percentage. As it turns out, mmag to percent depth conversions can be made simply by dividing the mmag measurement by ten (Still, 2013). However, despite the fact that we could not discern this transit in our data or freshen TOI 717.01's midpoint, we were able to make some interesting observations about other stars in its field, as will be discussed below.

## 3. Instruments Used

The Las Cumbres Observatory 0.4-m telescopes were used to obtain images of the targets. All of these telescopes have the same specifications: the aperture is 0.4m, the focal length is 3251mm (f/8), and the field of view is 29.2 x 19.5 arcminutes. All images were taken using a Pan-STARRS w-filter. The SBIG STL-6303 cameras have a 0.571 arcsecond pixel scale, and a 14-second overhead per frame. Because of the identical specs, the images were directly comparable even when they came from different observatories. Specifically, for the full HAT-P-3 dataset and the first TOI 717.01 dataset, images were returned from the Teide Observatory in Tenerife, Spain. For the second TOI 717.01 dataset, images were returned from Cerro Tololo in Chili. For the third TOI 717.01 dataset, images were returned from Haleakala in Maui. Eight separate Qatar-8 b requests were made, but all of these expired, meaning that the images could not be taken during the specified time windows. This can happen due to weather, telescope maintenance issues, or other astronomers' conflicting image requests.

## 4. Methods

Image requests were facilitated by Michael Fitzgerald's ExoRequest script (Sarva et. al., 2019). This script identifies the time to schedule an image request by calculating the predicted transit midpoint in conjunction with the epoch of the transit and the time prior to ingress and after egress for which baseline images are desired. In all cases, a window of 60 minutes on either side of the expected midpoint was requested. The script uses the host star's magnitude to identify an appropriate exposure time. The exposure times for TOI 717.01, HAT-P-3, and Qatar-8 b were 68 seconds, 24 seconds, and 26 seconds respectively.

The images received from the LCOGT were prereduced using the Our Solar Siblings (OSS) pipeline (Fitzgerald, 2018). In the OSS pipeline, six photometric algorithms are used to find the brightness's of stars in an image from the brightness's of their constituent pixels. The six algorithms include three types of aperture photometry: Aperture Photometry Tool (APT),

| Image | EXOTIC Mid-transit (BJD - 245000) | Mid-transit Uncertainty (days) | Difference From Expected Mid-transit (minutes) |
|---|---|---|---|
| (a) | 8890.35698 | 0.025408 | 10.2 |
| (b) | 8890.36797 | 0.001271 | 26.0 |
| (c) | 8903.56677 | 0.033836 | 76.1 |
| (d) | 8903.59338 | 0.013823 | 37.8 |

**Table 2. TOI 717.01 Mid-transit times.**

Source Extractor (SEX), Source Extractor Kron (SEK). In addition, they include three types of pointspread function photometric algorithms: DAOphot photometry (DAO), DOPhot photometry (DOP), and PSFEx photometry (PSX). Details of the algorithms appear in the references below (Bertin, 2011; Bertin and Arnouts, 2018; Laher et al., 2012, Schechter and Mateo, 1993; Stetson, 1987).

## 5. Observations of TOI 717.01

Using the LCOGT images and the EXOplanet Transit Interpretation Code (EXOTIC) (Zellem et al., 2020), light curves of TOI 717.01 were generated using several different comparison stars (comp stars). Figure 3 (at end of paper) shows the lightcurves from EXOTIC using different comp stars on the three datasets, and Table 2 shows the corresponding fitted mid-transit times with their uncertainties. In Figure 3, (a) comes from images taken at Teide on February 11th, 2020, (b) and (c) come from images taken at Cerro Tololo on February 24th, 2020, and (d) was from images taken on Haleakala on March 9th, 2020. The difference between the observed transit midpoint and that listed on the NASA Exoplanet archive was divided by the period to compute the number of periods that had passed since the listed transit midpoint. The decimal part of the number resulting from that operation was the difference in phase between the observed and expected transit midpoint; this was converted into minutes for inclusion in Table 2. The changes in position and depth of the fitted transit between these sets of images confirm that the transit of TOI 717.01 causes a smaller dip than can be detected using the LCO equipment.

Table 3 (at end of paper) shows the comp stars used for differential photometry of TOI 717.0. The brightness variation over the course of the Teide images, which are labeled (a) in Figure 3, may be evidence of variability in the corresponding comp star, even though the comp is not marked as a variable star in the Gaia DR2 or SIMBAD databases.

The comp star used in the Haleakala images, which are labeled (d) in Figure 3, was a red giant, so the slight protuberance near phase 0.05 of (d) might be evidence of stellar activity in the comp corresponding to its stage of life.

## 6. Observations of HAT-P-3b

Lightcurves of the HAT-P-3 b system were constructed using EXOTIC. In Figure 4 (a) and (b) (at end of paper), which used EXOTIC version 0.6.5, the transit appears to be in process at the beginning of the fit, even though images in the series included a one-hour window before and after the transit. However, after removing certain faulty images and running them through EXOTIC version 0.6.9, we obtained the light curve seen in Figure 4 (c) which shows a clean transit with a clear dip. The faulty images that were removed included the first image, which contained a strange image header and had a different appearance from the other images. The 57$_{th}$ image in the series was also removed, because of a warning message from EXOTIC about a drifting comparison star in that image. It is likely that the updated software version and also the removal of the corrupted images contributed to the cleaner appearance of the light curve in Figure 4 (c). As seen in Table 4, EXOTIC reported a mid-transit time of 2458907.6298, corresponding to a phase difference of approximately 0.01645 (23.683 minutes).

As a comparison, EXOTIC was run on prereduced data from the six different LCO photometric reduction algorithms. The corresponding light curves are seen in Figure 5 and Table 4 (at end of paper). Values for the transit midpoint, depth, and uncertainty vary widely among different photometry types, with SEK photometry showing the lowest residual scatter and lowest transit depth uncertainty. However, running EXOTIC directly on the images produced lower residual scatter and transit depth uncertainty than any of the other the photometry types.

## 7. Conclusion

For HAT-P-3, we measured a transit midpoint of 2458907.6298 BJD using EXOTIC. Running EXOTIC directly on image files had a lower residual scatter than running it on pre-reduced data from other photometric algorithms. It is possible, but not confirmed, that a comparison star used for TOI 717.01 was variable. Our data for both HAT-P-3 b and TOI 717.01 have been uploaded to the American Association of Variable Star Observers (AAVSO) website.

## 8. Acknowledgements

This study has made use of the NASA Exoplanet Archive, which is operated by the California Institute of Technology, under contract with the National Aeronautics and Space Administration under the Exoplanet Exploration Program. This paper includes data collected by the Kepler and K2 mission. Funding for the Kepler and K2 mission is provided by the NASA Science Mission directorate.

This research used data from the European Space Agency's Gaia Data Release 2 (https://www.cosmos.esa.int/gaia), processed by the Gaia Data Processing and Analysis Consortium (DPAC, https://www.cosmos.esa.int/web/gaia/dpac/consortium). Funding for the DPAC has been provided by national institutions, in particular, those participating in the Gaia Multilateral Agreement.

In addition, the authors would like to thank the Las Cumbres Observatory Global Telescope Network for the telescope time and the Our Solar Siblings Pipeline (OSS-Pipeline for photometric reduction.

This publication makes use of data products from Exoplanet Watch, a citizen science project managed by NASA's Jet Propulsion Laboratory on behalf of NASA's Universe of Learning. This work is supported by NASA under award number NNX16AC65A to the Space Telescope Science Institute.

## 9. References

Bertin, E. "Automated Morphometry with SExtractor and PSFEx." (2011). *Astronomical Data Analysis Software and Systems XX*. ASP Conference Series, Astronomical Society of the Pacific, **442**, pp. 435 – 438.

Bertin, E., & Arnouts, S. "Sextractor: Software for source extraction" (2013) *Astronomy and Astrophysics Supplement Series*, **117**, 393–404.

Brown, T., Baliber, N., Bianco, F., et al, "Las Cumbres Observatory Global Telescope Network" (2013) *Pub. Astron. Soc. Pac.*, **125**,1031–1055.

Fitzgerald, M. "The Our Solar Siblings Pipeline: Tackling the data issues of the scaling problem for robotic telescope based astronomy education projects. Robotic Telescopes" (2018). *Student Research, and Education Proceedings*, **1**.

Laher, Russ R. et al. "Aperture Photometry Tool." (2012). *Pub. Astron. Soc. Pac.*, **124**, 737-763. ISSN 0004-6280, 1538-3873.

Sarva, Jay, et al. "An Exoplanet Transit Observing Method using LCO Telescopes, ExoRequest and Astrosource" (2020). *Astronomy Theory, Observations, and Methods*, **1**, Accepted.

Schechter, P., & Mateo, M. "DOPhot, a CCD photometry program: Description and tests" (1993) *Pub. Astron. Soc. Pac.*,**105**, 1342–1353.

Stetson, P. "Daophot: A computer program for crowded-field stellar photometry" (1987). *Pub. Astron. Soc. Pac.*, **99**, 191–222.

Still, M. "Kepler Calibration: Signal-to-Noise Characteristics" (2013). *National Aeronautics and Space Administration website*. Edited by Martin Still. Retrieved April 2020.

Zellem, Robert et al., "Utilizing Small Telescopes Operated by Citizen Scientists for Transiting Exoplanet Follow-up" (2020). *Pub. Astron. Soc. Pac.*, **132**, id.054401 10.1088/1538-3873/ab7ee7

| Target | Orbital Period (Days) | Transit Depth | Transit Midpoint (BJD) | Transit Duration (hours) | Apparent Magnitude | Stellar Temperature |
|---|---|---|---|---|---|---|
| TOI 717.01 | 0.5079 | 1 mmag | 2458518.26611 | 1.238 | 11.44 | 3427 |
| Qatar-8b | 3.715 | 1.01% | 2458210.83890 | 4.027 | 11.34 | 5738 |
| HAT-P-3b | 2.899 | 1.24% | 2454856.70118 | 2.075 | 11.58 | 3185 |

**Table 1. Planetary, stellar, and transit properties of exoplanet target candidates.**

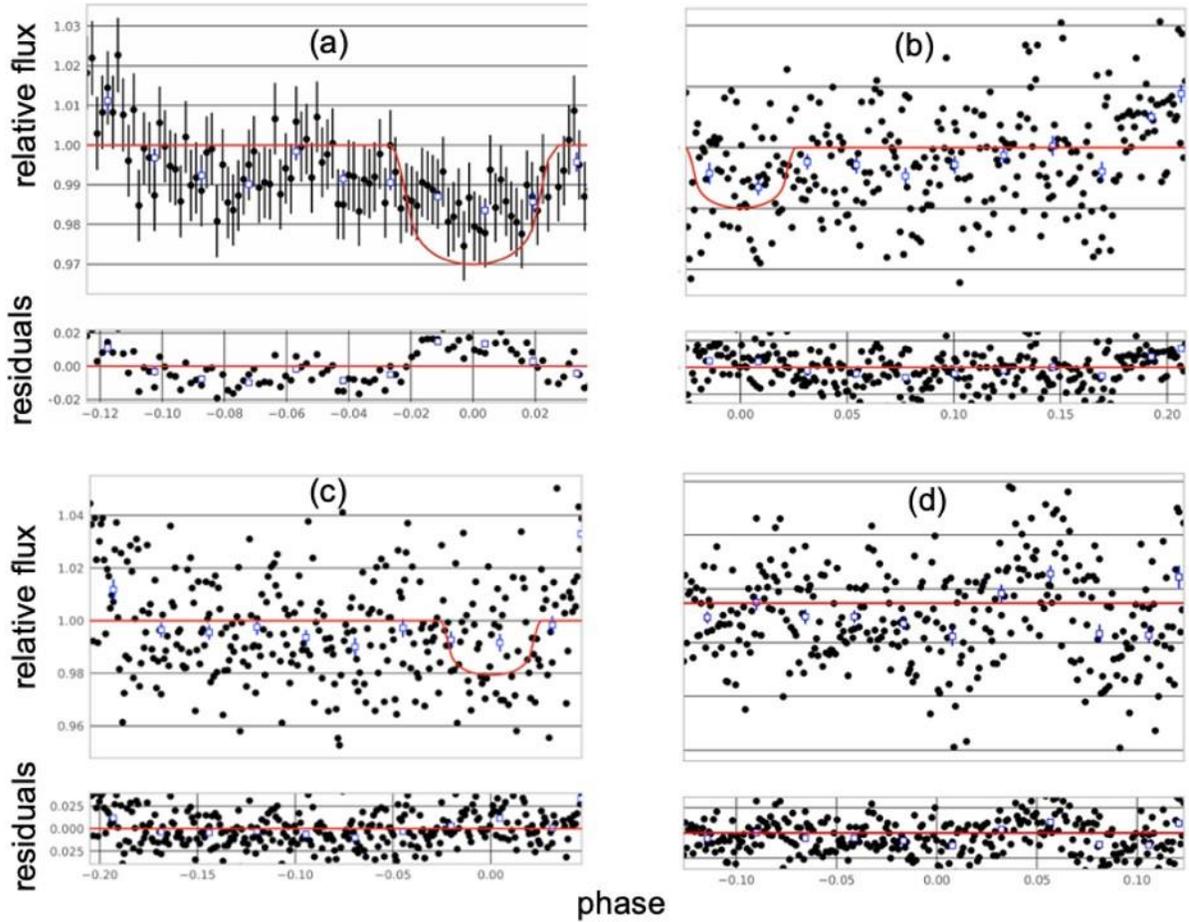

**Figure 3.** Light curves of the TOI 717.01 system using EXOTIC directly on FITS files received from LCOGT, with image (a) from the first set of data, images (b) and (c) from the second set of TOI data, and image (d) from the third set of data.

| Lightcurve | Comp RA | Comp Dec | Apparent Mag | Parallax Angle (mas) | Absolute Magnitude | Red Giant |
|---|---|---|---|---|---|---|
| (a) | 09:51:57 | 02:07:01 | 12.74 | 28.76 | 10.03 | No |
| (b) | 09:51:57 | 02:07:01 | 12.74 | 28.76 | 10.03 | No |
| (c) | 09:51:55 | 02:09:16 | 13.67 | 11.01 | 8.88 | No |
| (d) | 09:51:27 | 02:05:20 | 11.84 | 0.64 | 0.86 | Yes |

**Table 3.** Comparison Star information regarding Figure 3.

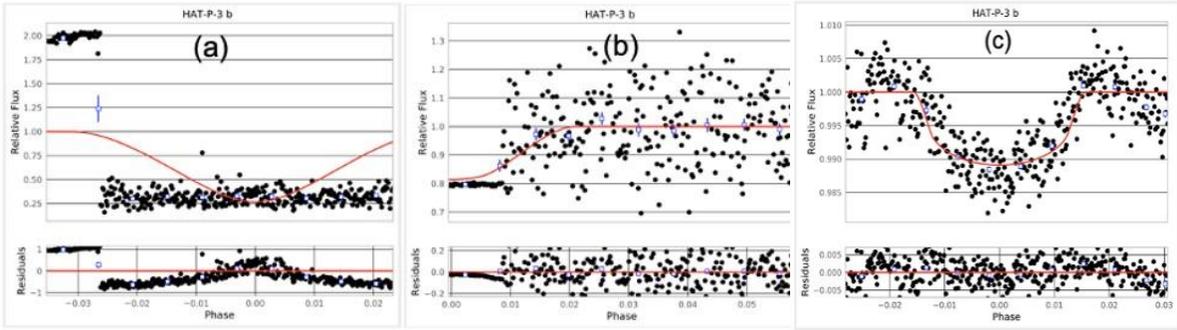

**Figure 5.** EXOTIC light curves of HAT-P-3 b using FITS files.

| Reduction method | Transit Depth | Depth Uncertainty | Transit Midpoint (BJD) | Midpoint Uncertainty (Days) | Scatter in Residuals (%) |
|---|---|---|---|---|---|
| Fits Files | 0.911 | 0.059 | 2458907.6298 | 0.001447 | 0.354 |
| PSX | 2.15 | 0.294 | 2458907.6267 | 0.002886 | 0.774 |
| SEX | 1.126 | 0.115 | 2458907.6268 | 0.002876 | 0.446 |
| SEK | 0.972 | 0.109 | 2458907.6205 | 0.003552 | 0.51 |
| APT | 2.14 | 0.2584 | 2458907.6199 | 0.004547 | 1.169 |
| DAO | 3.617 | 1.246 | 2458907.5456 | 0.0572 | 2.115 |
| DOP | 3.8396 | 1.256 | 2458907.5699 | 0.009163 | 7.324 |

**Table 4.** Information on transit properties for the HAT-P-3 light curves.

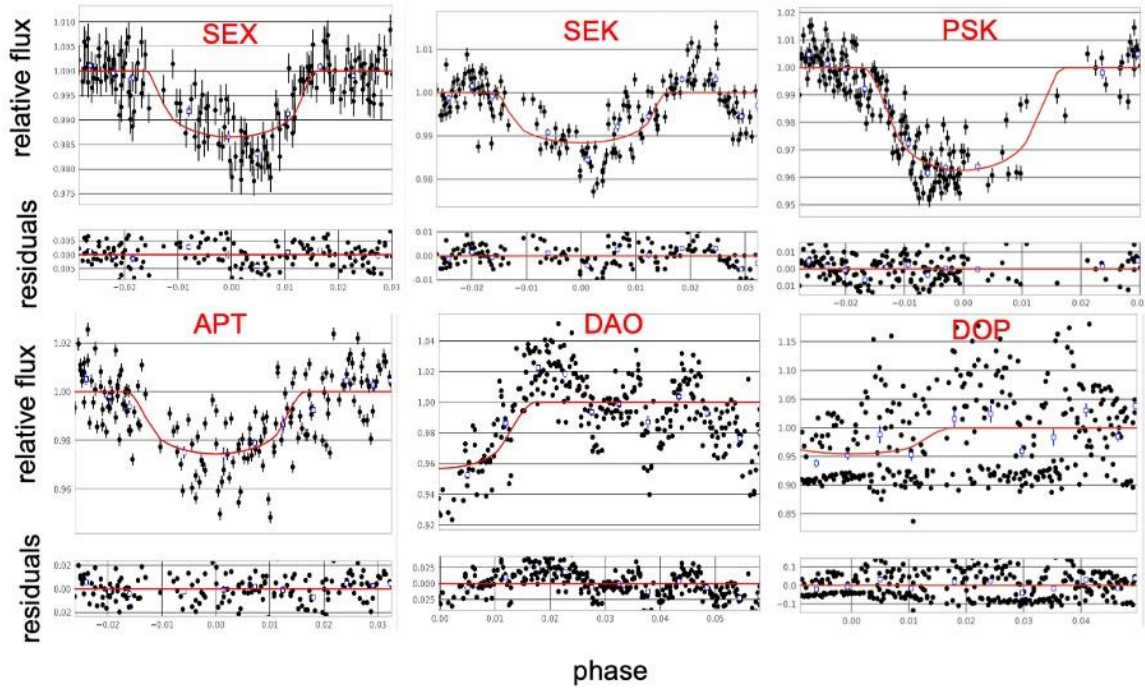

**Figure 4. Light curves of HAT-P-3 b using the SEX (a), SEK (b), PSX (c), APT (d), DAO (e). and DOP (f) photometry. The graphs in conjunction with Table2depict that the SEK and SEX photometry types generally had lower uncertainties and were closer to the expected transit depth.**